\documentclass[pre,twocolumn]{revtex4-2}
\usepackage{graphicx}
\usepackage{amsmath,amssymb}

\newcommand{\Tr}{{\rm Tr}}

\newcommand{\be}{\begin{equation}}
\newcommand{\ee}{\end{equation}}

\begin{document}

\title{Effect of a tunnel barrier on time delay statistics}
\author{Marcel Novaes}
\affiliation{Instituto de F\'isica, Universidade Federal de Uberl\^andia, Uberl\^andia, MG, 38408-100, Brazil}
\author{Jack Kuipers}
\affiliation{D-BSSE, ETH Zurich, Mattenstrasse 26, 4058 Basel, Switzerland}

\begin{abstract}
We develop a semiclassical approach for the statistics of the time delay in quantum chaotic systems in the presence of a tunnel barrier, for broken time-reversal symmetry. Results are obtained as asymptotic series in powers of the reflectivity of the barrier, with coefficients that are rational functions of the channel number. Exact expressions, valid for arbitrary reflectivity and channel number, are conjectured and numerically verified for specific families of statistical moments. 
\end{abstract}

\maketitle

\section{Introduction}

We consider the problem of quantum scattering in complex systems such as a cavity with chaotic dynamics, which is connected to the outside by a finite number of scattering channels, $M$. The amplitudes of outgoing waves are given in terms of the incoming waves by multiplication with the $M$-dimensional $S$ matrix, which is unitary when there is no dissipation. We assume that, at any given energy, the classical dynamics is characterized by a well defined dwell time $\tau_D$, the average time spent in the scattering region. We also assume broken time-reversal symmetry.

The energy derivative of the logarithm of the scattering matrix, $S$, is known as the Wigner–Smith matrix \cite{time1,time2,time3,time4},
\be Q=-i\hbar S^\dagger \frac{dS}{dE},\ee
where $\hbar$ is Planck's constant. This is an operator representing the quantization of the notion of time delay, i.e.\ the time spent by quantum particles inside the scattering region. Its normalized trace $\tau_W = \frac{1}{M}\Tr(Q)$ is called the Wigner time delay. 

For complex systems, the matrix elements of $Q$ are typically widely fluctating functions of the energy and it is advisable to restrict attention to local averages of observables. The local average of the Wigner time delay at a given energy, for example, equals precisely the classical dwell time at that energy, $\langle\tau_W \rangle=\tau_D$. More refined statistical information about the time duration of wave scattering is encoded in other spectral properties of  $Q$.

Central to our approach is the Schur polynomial $s_\lambda(Q)$, a symmetric polynomial of the eigenvalues of $Q$ which is defined in terms of an integer partition $\lambda=(\lambda_1,\ldots,\lambda_{\ell(\lambda)})$, i.e.\ a non-decreasing sequence of $\ell(\lambda)$ positive integers. Of crucial importance is the fact tha every symmetric polynomial in $Q$ can be written as a linear combination of Schurs. For example, $\tau_W=s_{(1)}(Q)$ and $\tau_W^2=\frac{1}{2}s_{(2)}(Q)+\frac{1}{2}s_{(1,1)}(Q)$. We call the local averages $\langle s_\lambda(Q)\rangle$ the Schur-moments of $Q$.

Within a random matrix theory (RMT) approach, detailed characteristics of the system are left aside and $Q$ is instead treated as a random matrix \cite{rmt0,rmt1,rmt2,rmt3,rmt4}. This is a fruitful point of view that has lead to many interesting results \cite{results1,results2,results3,results4,cunden,results5,results6,results7,results8,results9,
eu1,grabsch,eu2}. In particular, when time-reversal symmetry is not present, which is the case we consider here, an explicit expression can be found for Schur-moments \cite{eu1},
\be\label{rmt} \langle s_\lambda(Q)\rangle=(M\tau_D)^{|\lambda|}\frac{d_\lambda [M]^\lambda}{|\lambda|![M]_\lambda},\ee
in terms of quantities we define later.

We address in this work the effect on the time delay statistics of introducing an imperfect coupling between the scattering region and the exterior, such as a tunnel barrier of reflection probability $\gamma$. No effect at all exists on the average time delay, $\langle s_{(1)}(Q)\rangle(\gamma)=\langle s_{(1)}(Q)\rangle(0)$, and this can be understood semiclassically as follows: in the presence of the barrier, an incident particle may be reflected promptly without delay, with probability $\gamma$, or enter the cavity with probability $1-\gamma$; after a time $\tau_D$, it tries to leave and succeeds with probability $1-\gamma$ or is reflected back inside with probability $\gamma$; and so on. Summing over all possibilities leads to a total average time delay which is
\be (1-\gamma)\left(\tau_D+2\tau_D\gamma+3\tau_D\gamma^2+\cdots\right)(1-\gamma)=\tau_D.\ee However, the presence of the barrier influences higher statistics, so that in general $\langle s_{\lambda}(Q)\rangle(\gamma)\neq \langle s_{\lambda}(Q)\rangle(0)$. For example, we find that
\be\label{t2} \langle \tau_W^2\rangle=\tau_D^2+\frac{2\tau_D^2}{(1-\gamma)^2(M^2-1)}.\ee In particular, the variance of $\tau_W$ becomes infinite when $\gamma\to 1$ at finite $M$, but may attain any value if $M$ scales as $(1-\gamma)^{-1}$.

Tunnel barriers have been considered before in the context of time delay \cite{rmt0,rmt2}, but always under some approximation. For instance, the number of channels is either taken to be very small, like one \cite{gopar} or two \cite{results8}, or instead it is taken to be very large, $M\gg 1$ \cite{rmt2,results7}. This is moreover compounded with the assumption that the reflection probability is either very small or very close to one \cite{grabsch}. In contrast, our results are valid for arbitrary values of both these parameters, as can be appreciated from Eq.(\ref{t2}).

We rely on the semiclassical approach which has proven very successful in quantum chaos \cite{semi1,semi2,semi3,delay0,delay1,delay2,delay3}. Specifically we develop a combination of the approach introduced by Kuipers, Savin and Sieber for the time delay \cite{efficient} and the formulation in terms of matrix integrals \cite{matrix1,matrix2,matrix3}.

The work is organized as follows. In Section \ref{sec:semi} we introduce the semiclassical approach that leads to an infinite series in powers of $\gamma$ for $\langle s_{\lambda}(Q)\rangle(\gamma)$, the coefficients of which are rational functions of $M$. In Section \ref{sec:conj} we conjecture closed forms for some of these functions and present some special cases. In particular, our results imply that for $M(1-\gamma)\gg 1$ the distribution of $\tau_W$ tends to a Gaussian of mean $\tau_D$ and variance $\frac{2\tau_D^2}{M^2(1-\gamma)^2}$. In Section \ref{sec:numer} we validate our conjectures against simulations of a concrete system, a chaotic quantum map. We conclude in Section \ref{sec:conc}. For simplicity in the following we now set $\tau_D=1$ and measure all times in units of $\tau_D$.

\section{Semiclassical approach} \label{sec:semi}

\subsection{Diagrammatic rules} \label{sec:semi_diag}

The first diagrammatical semiclassical approaches to the time delay relied on energy correlations of the scattering matrix to derive their results \cite{kuipers,kuipersrichter,novaessemi,euzinho}. In \cite{efficient}, a more efficient semiclassical approach to time delay was developed in such a way that the calculation of average traces $\langle\Tr(Q^n)\rangle$ requires
$n$ trajectories $\sigma$ and $n$ trajectories $\sigma'$ to enter the chaotic region, with $\sigma_k$ going from scattering channel $i_k$ to a certain end point $r_k$ inside the cavity, and $\sigma_k'$ going from channel $i_{k+1}$ to the same end point $r_k$. These two sets of trajectories must be correlated, in the sense that they have approximately the same total action and actually differ from each other only in the vicinity of a structure called ``encounters'' which have been recognized \cite{semi1,semi2} to be the mechanism responsible for systematic constructive interference between sets of trajectories. 

A diagrammatic perturbative theory for time delay moments then follows with diagrams consisting of initial and final vertices corresponding to channels and endpoints, together with internal vertices corresponding to encounters. These vertices are connected by edges, corresponding to long stretches of chaotic motion. The contribution of a given diagram contains several multiplicative factors: $M$ for each channel, $1/M$ for each edge and $-M$ for each vertex that does not contain an end point; vertices that contain one end point contribute a factor $1$ and those with more than one end point contribute $0$. 

These diagrammatic rules were used in \cite{PD} to obtain all Schur-moments of the time delay, in complete agreement with the RMT prediction in Eq.~(\ref{rmt}). Here we extend this model to account for the presence of the tunnel barrier. The contribution of a given diagram is still made up of multiplicative factors but, if the reflection probability is $\gamma$, then they are $M(1-\gamma)$ for each channel, $[M(1-\gamma)]^{-1}$ for each edge, $-M(1-\gamma^q)$ for each vertex of valence $2q$ that does not contain an end point nor happen at the lead, $\gamma$ for each time an edge reflects off the barrier. Vertices with one end point still contribute $1$ and those with more than one end point still contribute $0$. 

We show an example in Figure 1, a diagram that contributes to the semiclassical calculation of $\langle({\rm Tr}(Q))^2\rangle$. The barrier is represented by the shaded rectangle, trajectories $\sigma$ by solid lines and $\sigma'$ by dashed ones. The end points are marked A and B. There is an encounter inside the cavity, grossly magnified for effect, labelled C. The contribution of C is $-M(1-\gamma^2)$ if it happens far enough from B, otherwise it is 1. Encounter D happens at the lead with one reflection, so its contribution is $\gamma$. There are five edges, each contributing $[M(1-\gamma)]^{-1}$. Transmission through the barrier gives $(1-\gamma)^2$, so the total contribution of this diagram, when C and B are far apart, is 
\be -\frac{(1-\gamma)^2\gamma M(1-\gamma^2)}{M^5(1-\gamma)^5}=-\frac{\gamma (1-\gamma^2)}{M^4(1-\gamma)^3}.\ee

We implement such rules by means of an appropriately designed matrix integral, a method that was introduced in \cite{matrix1,matrix2,matrix3} and applied to several different situations. Before introducing that integral, let us revise some background concepts.

\begin{figure}[t]
\centering
\includegraphics[scale=1.3]{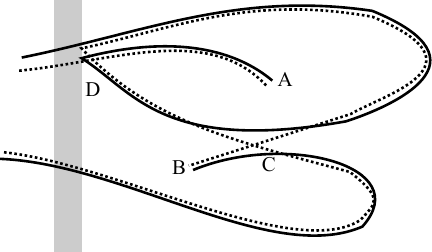}
\caption{A diagram that contributes to the semiclassical calculation of $\langle({\rm Tr}(Q))^2\rangle$. The barrier is represented by the shaded rectangle, trajectories $\sigma$ by solid lines and $\sigma'$ by dashed ones. There is one encounter inside the cavity, marked C, while encounter marked D happens at the lead. A and B are endpoints.}
\end{figure}

\subsection{Permutation groups and unitary groups}

In this subsection we collect several basic facts from combinatorics and group theory that are needed for our calculations. The reader who is familiar with these concepts may skip this. Standard references for this material are \cite{mac,stan}.

The power sum symmetric polynomials in $Q$ are products of traces of powers of $Q$:
\be p_\mu(Q)=\prod_{i=1}^{\ell(\mu)}\Tr(Q^{\mu_i}),\ee
where $\mu=(\mu_1,\ldots,\mu_{\ell(\mu)})$ is an integer partition. We write $\mu\vdash n$ or $|\mu|=n$ to denote the fact that $\sum_i \mu_i=n$. Like the Schur polynomials, power sums also provide a basis for the vector space of symmetric polynonials in the eigenvalues of $Q$. The transition between these basis is made by the matrix of irreducible characters of the permutation group $S_n$, namely for any $\mu\vdash n$ we have
\be\label{decomp} p_{\mu}(Q)=\sum_{\lambda\vdash n}\chi_\lambda(\mu)s_\lambda(Q),\ee
where the characters $\chi$ satisfy the orthogonality relation
\be\label{orth} \frac{1}{n!}\sum_{\beta \in S_n}\chi_\lambda(\alpha\beta)\chi_\mu(\pi\beta)=\frac{\chi_\lambda(\alpha\pi^{-1})}{d_\lambda}\delta_{\lambda,\mu},\ee
where $\alpha,\beta,\pi$ are permutations and $d_\lambda=\chi_\lambda(1)$
is the dimension of the irreducible representation of $S_n$ labelled by $\lambda$.

Schur polynomials can be used in infinite series as well, such as 
\be \label{Cauchy}\frac{1}{\prod_{ij}(1-x_iy_j)}=\sum_\lambda s_\lambda(X)s_\lambda(Y),\ee
where we view $x_1,x_2,...$ as eigenvalues of $X$ and $y_1,y_2,...$ as eigenvalues of $Y$. The above relation, a generalization of the geometric series, is known as the Cauchy identity. There is also the Littlewood identity, 
\be\label{Littlewood}\frac{1}{\prod_{j>i}(1-x_ix_j)}=\det(1-X)\sum_\lambda s_\lambda(X).\ee

The Schur polynomials in $N$ variables have another important property: they are the characters of irreducible polynomial representations of the unitary group $\mathcal{U}(N)$. This means that
\be \int_{\mathcal{U}(N)}s_\lambda(U^\dagger)s_\mu(U)dU=\delta_{\lambda\mu},\ee
where $dU$ is the normalized Haar measure. These polynomials have a determinantal formula 
\be\label{det} s_\lambda(U)=\frac{\det(z_i^{N+\lambda_j-j})}{\Delta(U)},\ee
where $z_1,...,z_N$ are the eigenvalues of $U$ and 
\be\label{vdmonde} \Delta(U)=\det(z_i^{N-j})=\prod_{j<k}(z_k-z_j)\ee
is called the Vandermonde of $U$. Using this formula and the Andr\'eief integration formula \cite{andre}
\be \int d\vec{z}\det(f_i(z_j))\det(g_k(z_j))=N!\det\left(\int dz f_i(z)g_k(z)\right),\nonumber\ee
the orthogonality (\ref{orth}) can be proved.

The orthogonality of course allows Schur expansions can be computed, i.e.\ given some symmetric function $f(X)$, the coefficients $A_\lambda$ in $f(X)=\sum_\lambda A_\lambda s_\lambda(X)$ are given by 
\be\label{expan} A_\lambda= \int_{\mathcal{U}(N)} f(U)s_\lambda(U^\dagger)dU.\ee In particular, the coefficients in $s_\lambda(X) s_\mu(X)=\sum_\nu C^\nu_{\lambda\mu}s_\nu(X)$ are called the Littlewood-Richardson coefficients.

The last bit of information we need are the Weingarten functions. The average of a product of matrix elements from unitary matrices,
\be\label{weingU}\left\langle \prod_{t=1}^n U_{i_tj_t}U^*_{q_tp_t}\right\rangle_{\mathcal{U}(N)},\ee
is given by a sum over all possible permutations that may take $\vec{q}$ to $\vec{i}$ and $\vec{p}$ to $\vec{j}$, 
\be\sum_{\sigma,\tau \in S_n} \delta_\tau(\vec{q},\vec{i})\delta_\sigma(\vec{p},\vec{j}){\rm Wg}^{U}_N(\sigma^{-1}\tau), \ee where $\delta_\sigma(\vec{i},\vec{j})=\prod_{k=1}^n \delta_{i_k,j_{\sigma(k)}}$ and ${\rm Wg}^{U}_N$ are called the Weingarten functions. They have a simple character expansion \cite{weing},
\be {\rm Wg}^{U}_N(\pi)=n!\sum_{\substack{\lambda\vdash n\\\ell(\lambda)\le N}}\frac{d_\lambda \chi_\lambda(\pi)}{[N]^\lambda}.\ee

In the above formula we have a generalized rising factorial
\be [N]^\lambda=\frac{|\lambda|!s_{\lambda}(1_N)}{d_\lambda}=\prod_{i=1}^{\ell(\lambda)}\frac{(N+\lambda_i-i)!}{(N-i)!}. \ee
The corresponding generalized falling factorial is
\be [N]_\lambda=\prod_{i=1}^{\ell(\lambda)}\frac{(N+i-1)!}{(N-\lambda_i+i-1)!}. \ee
These are the quantities that appear in (\ref{rmt}). They satisfy two symmetry relations
\be \label{sym1}[-N]^\lambda=(-1)^{|\lambda|}[N]_\lambda\ee and 
\be \label{sym2}[N]^{\lambda'}=[N]_\lambda,\ee where $\lambda'$ is the partition conjugated to $\lambda$ (obtained by transposing its Ferrer's diagram).  
Clearly $[x]^\lambda$ is a polynomial in $x$ of degree $|\lambda|$. Let $t_\lambda$ be the coefficient of the smallest power of $x$ in this polynomial, so that, when $x$ is small, we have 
\be\label{t} [x]^\lambda=t_\lambda x^{D(\lambda)}+O(x^{D+1}),\ee with $D(\lambda)$ being the number of parts in $\lambda$ for which $\lambda_i-i\ge 0$.

\subsection{The matrix integral}

The matrix integral which encodes the semiclassical approach to the calculation of $\langle p_\mu(Q)\rangle$ is
\begin{multline} \sum_{\vec{i}}\frac{[M(1-\gamma)]^n}{\mathcal{Z}}\int dZe^{-M\sum_{q\ge 1}\frac{1-\gamma^q}{q}\Tr(Z^\dagger Z)^q}\\\times\prod_{k=1}^n \left(\frac{1}{1-Z^\dagger Z}Z^\dagger\right)_{i_k,k}\left(Z\frac{1}{1-\gamma Z^\dagger Z}\right)_{k,i_{\pi(k)}},\end{multline} where $Z$ is a complex $N$-dimensional matrix with a Gaussian distribution given by $e^{-M(1-\gamma)\Tr(Z^\dagger Z)}$. The normalization is $ \mathcal{Z}=\int e^{-M(1-\gamma)\Tr(Z^\dagger Z)}$. It bear some resemblance to the integrals used in \cite{tunnel,barriers} to compute transport moments in the presence of tunnel barriers. 

The integral is computed using Wick's rule \cite{wick1,wick2,wick3}, according to which each term like ${\rm Tr}(Z^\dagger Z)^q$ is a vertex of valence $2q$ and contractions between matrix elements of $Z$ are represented by edges. The term $e^{-M\sum_{q\ge 2}\frac{1}{q}\Tr(Z^\dagger Z)^q}$, when expanded in powers of $M$, produces all possible vertices without end points, the ones of valence $2q$ being accompanied by a factor $-M(1-\gamma^q)$. Likewise, the presence of $M(1-\gamma)$ in the Gaussian measure leads to the contribution of $[M(1-\gamma)]^{-1}$ for each edge. The terms therefore exactly match the semiclassical contributions in Section \ref{sec:semi_diag}.

Besides the exponential term, which is like an ``internal'' part, we have two more terms, which are like ``channel'' parts. The quantity $\left(\frac{1}{1-Z^\dagger Z}Z^\dagger\right)_{i_k,k}$ represents trajectories going from channel $i_k$ to end point $r_k$; the geometric series produces all possible vertices that contain one end point. Analogously, the quantity $\left(Z\frac{1}{1-\gamma Z^\dagger Z}\right)_{k,i_{\pi(k)}}$ represents trajectories going from channel $i_{\pi(k)}$ to end point $r_k$, with $\pi$ being any permutation with cycle type $\mu$; the geometric series now produces all encounters that happen at the lead, in which trajectories may be reflected (leading to a factor $\gamma$ for each reflection). This term already appeared in \cite{tunnel}. Finally, the sum over $i_1,\ldots,i_n$ takes into account all possible channels through which a trajectory may enter the chaotic region. On the other hand, the end points are labelled by a different integer and hence cannot be equal. 

The resulting diagrammatical formulation of this matrix integral indeed therefore coincides with the semiclassical rules we discussed in Section \ref{sec:semi_diag}. Except that we must exclude all diagrams that contain closed cycles, i.e. periodic orbits, since these are not be present in the semiclassical approach. Closed cycles give rise to powers of $N$: the contribution of a diagram with $t$ closed cycles is proportional to $N^t$. Therefore, we consider the part of the result that is constant with respect to $N$ or, equivalently, we take the limit $N\to 0$.

\subsection{The solution}

Introduce the singular value decomposition $Z=UDV$, with $U$ and $V$ in the unitary group $\mathcal{U}(N)$ and $D$ a real and non-negative diagonal matrix. The Jacobian of this transformation is $dZ=dUdVdX\Delta(X)^2$ with $X=D^2$ and $\Delta(X)$ being the Vandermonde (\ref{vdmonde}). The channel parts involve
\be\left\langle V_{i_ka_k}V^{\dagger}_{c_ki_k}\right\rangle_{\mathcal{U}(N)}
\left\langle U_{kc_k}U^{\dagger}_{a_k\pi(k)}\right\rangle_{\mathcal{U}(N)}\frac{D_{a_k}D_{c_k}}{(1-X_{a_k})(1-\gamma X_{c_k})}.\nonumber\ee
Using the Weingarten function and character orthogonality, this becomes, after summing over $\vec{i},\vec{a},\vec{c},$
\be\sum_\lambda \chi_\lambda(\pi) s_\lambda\left(\frac{X}{(1-X)(1-\gamma X)}\right)\frac{[M]^\lambda}{([N]^\lambda)^2}.\ee
The Schur function above has a very complicated argument, so we must expand it in terms of regular Schur functions. First, we write it explicitly using the determinantal formula (\ref{det}).
Then we use that
\be \Delta\left(\frac{X}{(1-X)(1-\gamma X)}\right)=\frac{\Delta(X)\prod_{j>i}(1-\gamma x_ix_j)}{\det[(1-X)(1-\gamma X)]^{N-1}},\ee
and, by the Littlewood identity (\ref{Littlewood})
\be \frac{1}{\prod_{j>i}(1-\gamma x_ix_j)}=\det(1-\sqrt{\gamma}X)\sum_\alpha s_\alpha(\sqrt{\gamma}X).\ee
So $s_\lambda\left(\frac{X}{(1-X)(1-\gamma X)}\right)$ equals $\det(1-\sqrt{\gamma}X)$ times
\be\sum_\alpha s_\alpha(\sqrt{\gamma}X)\frac{1}{\Delta(X)}\det\left(\frac{x_k^{N+\lambda_i-i}}{[(1-x_k)(1-\gamma x_k)]^{\lambda_i-i+1}}\right).\nonumber\ee

Now we expand the last two quantities as a linear combination of Schurs. This is done using the expansion formula (\ref{expan}) and the Andr\'eief integration formula. Hence, we need to compute
\be F_{\lambda\rho}=\oint \frac{z^{N+\lambda_i-i}}{[(1-z)(1-\gamma z)]^{\lambda_i-i+1}}\bar{z}^{N+\rho_j-j},
\ee
where the integral is around the unit circle in the complex plane. 
Expanding $[(1-z)(1-\gamma z)]^{-(\lambda_i-i+1)}$ as
\be \sum_{k_1k_2=0}^\infty\binom{\lambda_i-i+k_1}{\lambda_i-i}\binom{\lambda_i-i+k_2}{\lambda_i-i}\gamma^{k_2}z^{k_1+k_2}\ee
we get
\be F_{\lambda\rho}=\det\left[\sum_k\binom{\rho_j-j-k}{\lambda_i-i}\binom{\lambda_i-i+k}{\lambda_i-i}\gamma^{k}\right].\ee

So
$s_\lambda\left(\frac{X}{(1-X)(1-\gamma X)}\right)$ equals  $\det(1-\sqrt{\gamma}X)$ times
\be\left(\sum_\rho F_{\lambda\rho}s_\rho(X)\right)\left(\sum_\alpha \sqrt{\gamma}^{|\alpha|}s_\alpha(X)\right).\ee Combining products of Schurs as linear combinations of Schurs by means of the Littlewood-Richardson coefficients, we can write this as
\be\sum_{r,\alpha,\rho} \sum_{\mu\nu}C_{1^r,\rho}^\mu C_{\mu\alpha}^\nu (-\sqrt{\gamma})^r F_{\lambda\rho}\sqrt{\gamma}^{|\alpha|}s_\nu(X).\ee

\begin{widetext}
The integral over the eigenvalues of $Z$ is
\be\label{matrix}\lim_{N\to 0}\sum_{\lambda\vdash n}\chi_\lambda(\pi)\frac{[M]^\lambda}{([N]^\lambda)^2}\frac{(1-\gamma)^n}{\mathcal{Z}}\int \frac{\det(1-X)^M}{\det(1-\gamma X)^M}|\Delta(X)|^2s_\nu(X)dX.\ee
Using the Cauchy identity (\ref{Cauchy}), we get
\be \lim_{N\to 0}(1-\gamma)^n\sum_{\lambda\vdash n}\chi_\lambda(\pi)\frac{[M]^\lambda}{([N]^\lambda)^2}\sum_\omega s_{\omega}(\gamma)\sum_\theta \frac{C_{\omega\nu}^\theta}{\mathcal{Z}}\int dX\det(1-X)^M|\Delta(X)|^2s_\theta(X).\ee
This is an integral of the Selberg type \cite{selberg}, and the result is known to be given by
\be (1-\gamma)^n\sum_{\lambda\vdash n}\chi_\lambda(\pi)\frac{[M]^\lambda}{t_\lambda^2}\sum_{\theta,D(\theta)=D(\lambda)} s_{\theta/\nu}(1^M)\gamma^{|\theta|-|\nu|}\frac{d_\theta t_\theta^2}{|\theta|![M]^\theta},\ee
where $s_{\theta/\nu}=\sum_\omega s_\omega C^\theta_{\omega\nu}$ is called a skew-Schur polynomial, and we have taken the limit $\lim_{N\to 0}\dfrac{[N]^\theta}{[N]^\lambda}=\dfrac{t_\theta}{t_\lambda}$, with $t_\lambda$ and $D(\lambda)$ as in (\ref{t}).

We therefore arrive at
\be \langle p_\pi(Q)\rangle=(1-\gamma)^n\sum_{\lambda\vdash n}\chi_\lambda(\pi)\sum_{\rho\nu} F_{\lambda\rho}G_{\rho,\nu} \frac{[M]^\lambda}{t_\lambda^2}\sum_{\theta,D(\theta)=D(\lambda)} s_{\theta/\nu}(1^M)\gamma^{|\theta|-|\nu|}\frac{d_\theta t_\theta^2}{|\theta|![M]^\theta},\ee
or, equivalently,
\be\label{final} \langle s_\lambda(Q)\rangle=(1-\gamma)^n\frac{[M]^\lambda}{t_\lambda^2}\sum_{\rho\nu} F_{\lambda\rho}G_{\rho,\nu} \sum_{\theta,D(\theta)=D(\lambda)} s_{\theta/\nu}(1^M)\gamma^{|\theta|-|\nu|}\frac{d_\theta t_\theta^2}{|\theta|![M]^\theta},\ee
\end{widetext}
where
\be G_{\rho,\nu}=\sum_{r,\alpha,\mu} C_{1^r,\rho}^\mu C_{\mu\alpha}^\nu (-\sqrt{\gamma})^r\sqrt{\gamma}^{|\alpha|}.\ee

This is convenient to put in the computer and find the first few terms in a power series in $\gamma$. However, we have not been able to sum this series and find a closed explicit formula.

\subsection{Asymptotic nature}

In particular, Eq.~(\ref{final}) seems to have poles at all values of $M$ because of the $[M]^\theta$ term in the denominator. But these poles are spurious and disappear when all sums are performed. 

For example, let us consider the calculation of the simplest average, $\langle s_{(1)}\rangle$. If we restrict $|\theta|\le 2$, we get from Eq.(\ref{final})
\be \frac{M^2-2}{M^2-1}-\gamma^2+\frac{\gamma}{M^2-1}.\ee This expression has a pole at $M=1$, while its large-$M$ expansion is
\be 1-\gamma^2-\frac{1-\gamma}{M^2}+O(M^{-4}).\ee
If we go up to $|\theta|\le 3$, we get a different approximation, more cumbersome, with another pole at $M=2$, but whose large-$M$ expansion is 
\be 1-\gamma^3-\frac{3\gamma(1-\gamma)}{M^2}+O(M^{-4}).\ee 
If we go further to $|\theta|\le 4$, we get yet a different approximation, even more cumbersome, that now has a third pole at $M=3$, but whose large-$M$ expansion is 
\be 1-\gamma^4-\frac{6\gamma^2(1-\gamma)}{M^2}+O(M^{-4}).\ee 
It is clear that when infinitely many terms are taken into account, as $\gamma<1$ the partial sum will get arbitrarily close to $1$, in agreement with expectation. The infinite series in Eq.~(\ref{final}) should be seen as an asymptotic series. 

As discussed in Section \ref{sec:conj}, we conjecture that these series actually represent some simple rational functions of $M$.

\subsection{Hook partitions}

The simplest kind of partitions are the hooks, $\lambda=(n-k,1^k)$. They can be used to compute the traces, because
\be p_\mu(Q)=\sum_{\lambda}\chi_\lambda(\mu)s_\lambda(Q)\ee
and $\chi_\lambda(n)$ is different from zero if and only if $\lambda$ is a hook.

When $\lambda$ is a hook, $\theta$ must also be a hook. This means $\nu$ must be a hook and thus also $\rho$. Let $\theta=(T,1^t)$, $\nu=(A,1^a)$, $\rho=(R,1^r)$. Then \begin{align} s_{\theta/\nu}(1^M)&=h_{T-A}(1^M)e_{t-a}(1^M)\\&=\binom{M+T-A-1}{T-A}\binom{M}{t-a},\end{align}
and, writing $(M)^{(n)}$ for the rising factorial and $(M)_{(n)}$ for the falling factorial,
\be \frac{d_\theta t_\theta}{|\theta|![M]^\theta}=\frac{(T-1)!t!}{(T+t)(M-t)^{(T+t)}}.\ee

The function $G_{\rho,\nu}$ is different from zero if $|\nu|-|\rho|$ is even and $\rho_1$ is either $\nu_1$ or $\nu_1-1$. For example, if $\nu=(3,1,1,1)$ then $\rho$ must belong to $\{(3,1,1,1),(2,1,1),(3,1),(2)\}$. Then we conjecture that $G_{\rho,\nu}=\gamma^{(|\nu|-|\rho|)/2}$.

\begin{widetext}
So we end up with the simpler expression
\be \frac{\langle s_{(n-k,1^k)}(Q)\rangle}{(1-\gamma)^n}=\sum_{\nu}\sum_\rho F_{\lambda\rho}\gamma^{(|\nu|-|\rho|)/2} \frac{[(M-k)^{(n)}]^2}{(n-k-1)!^2k!^2}\sum_{T,t,A,a} \binom{M+T-A-1}{T-A}\binom{M}{t-a}\frac{\gamma^{T+t-A-a}(T-1)!t!}{(T+t)(M-t)^{(T+t)}}.\ee
Unfortunately, it is still not so simple that would allow for explicit summation.
\end{widetext}

\section{Conjectures}\label{sec:conj}

Entering our expression into a computer algebra system and doing some ``experimental'' investigations, we have arrived at some conjectures.  

Inspection of Eq.(\ref{rmt}) shows that the symmetry relations (\ref{sym1}) and (\ref{sym2}) imply $\langle s_{\lambda'}\rangle(0,M)=(-1)^{|\lambda|}\langle s_\lambda\rangle(0,-M)$. Does this symmetry still hold in the presence of the tunnel barrier? We conjecture the answer to be yes:
\be\label{conj} \langle s_{\lambda'}\rangle(\gamma,M)=(-1)^{|\lambda|}\langle s_\lambda\rangle(\gamma,-M).\ee
We also conjecture another symmetry relation, involving reciprocity in the reflection probability:
\be \langle s_{\lambda'}\rangle(\gamma^{-1},M)[M]^{\lambda}=\langle s_\lambda\rangle(\gamma,M)[M]_{\lambda}.\ee

These symmetry relations suggest that maybe self-conjugate partitions ($\lambda=\lambda'$) should be particularly simple. Indeed, and quite surprisingly, we conjecture that for self-conjugate partitions the Schur-moment $\langle s_{\lambda}\rangle$ is actually independent of $\gamma$ and proportional to $M^{|\lambda|}$. For example, $\langle s_{(2,1)}\rangle(\gamma,M)=M^3/3$ and $\langle s_{(2,2)}\rangle(\gamma,M)=M^4/12$. So the fact that $\frac{1}{M}\langle {\rm Tr}(Q)\rangle$ does not dependent on $\gamma$ and $M$ generalizes to this whole class of Schur-moments.

Investigating the $\gamma$-series of Schur-moments, we have found evidence that, if these quantities are first multiplied by $(1-\gamma)^{|\lambda|}$, the series in fact terminates. So we conjecture that
\be (1-\gamma)^{|\lambda|}\langle s_{\lambda}\rangle(\gamma)=T_\lambda(\gamma),\ee
where $T_\lambda(\gamma)$ is a polynomial in $\gamma$. In the special case of singletons, $\lambda=(n)$, we conjecture the explicit form of this polynomial:
\be T_{(n)}(\gamma)=\frac{M^{n}}{n!(M)_{(n)}}\sum_{k=0}^n (-\gamma)^k\binom{n}{k}(M+k)^{(n-k)}(M-n+k)_{(k)},\ee
where $(x)^{(n)}$ and  $(x)_{(n)}$ are the usual rising and falling factorials.
Of course, a formula for  $T_{(1,...,1)}(\gamma)$ then follows from (\ref{conj}). For example,
\begin{align*}
\frac{T_{(2)}(\gamma)}{M^2}&=\frac{M(M+1)-2(M^2-1)\gamma+M(M-1)\gamma^2}{2M(M-1)},\\
\frac{T_{(1,1)}(\gamma)}{M^2}&=\frac{M(M-1)-2(M^2-1)\gamma+M(M+1)\gamma^2}{2M(M+1)}.
\end{align*}

Here is a more generic example, showing that even hooks can be complicated:
\begin{multline} 8M^{-4}(M-2)^{(4)}T_{(3,1)}(\gamma)=(M-1)^{(4)}\\-4(M-1)^2(M+1)(M+2)\gamma+6(M^2-4)(M^2+1/3)\gamma^2\\-4(M-1)(M-2)(M+1)^2\gamma^3+(M-2)^{(4)}\gamma^4.\end{multline}
The binomial numbers are still there, but a close look at the coefficient of $\gamma^2$ reveals that the dependence on $M$ may not factorize very nicely. 

\begin{figure*}[t]
\centering
\includegraphics[scale=0.5]{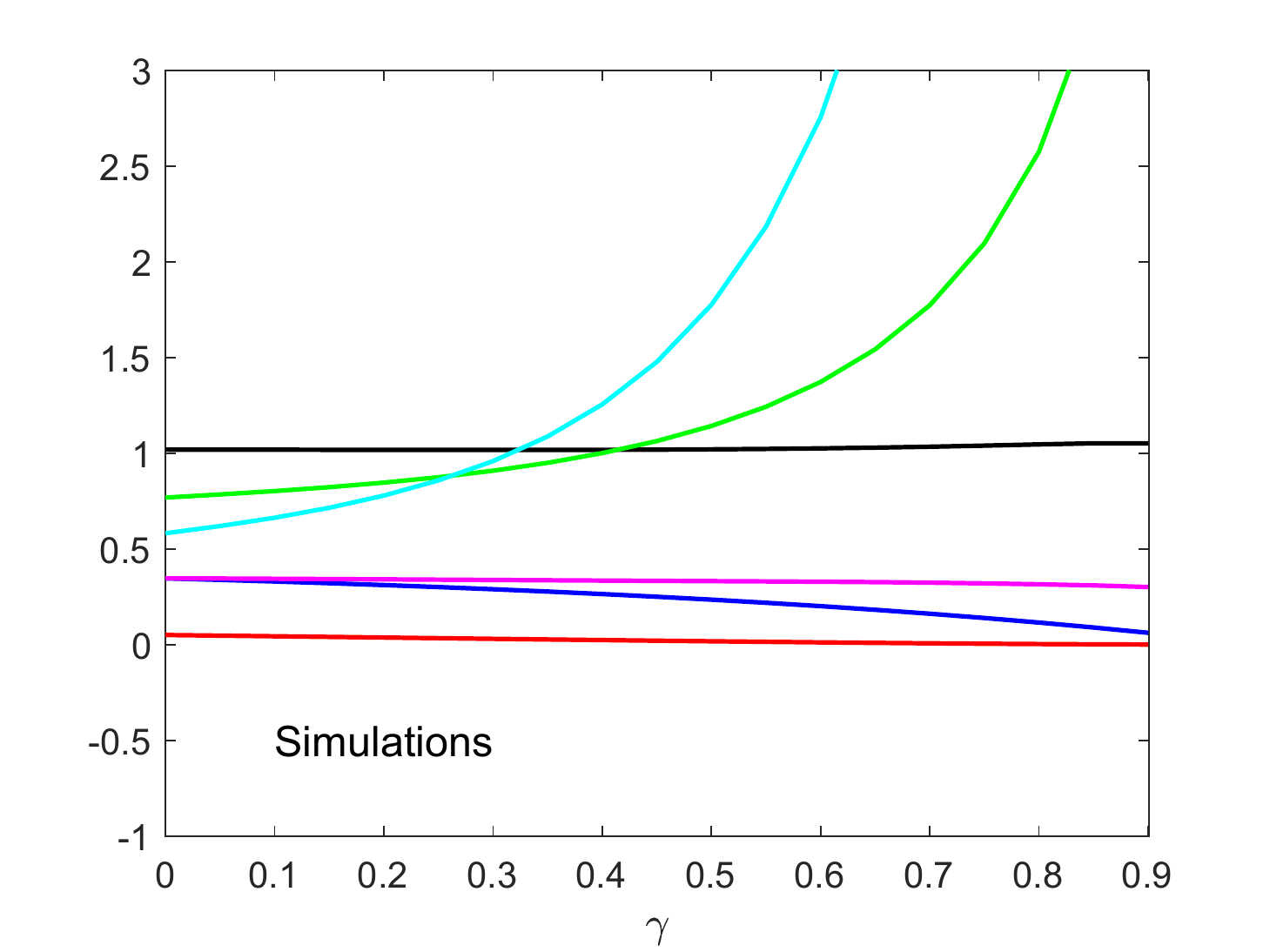}\includegraphics[scale=0.5]{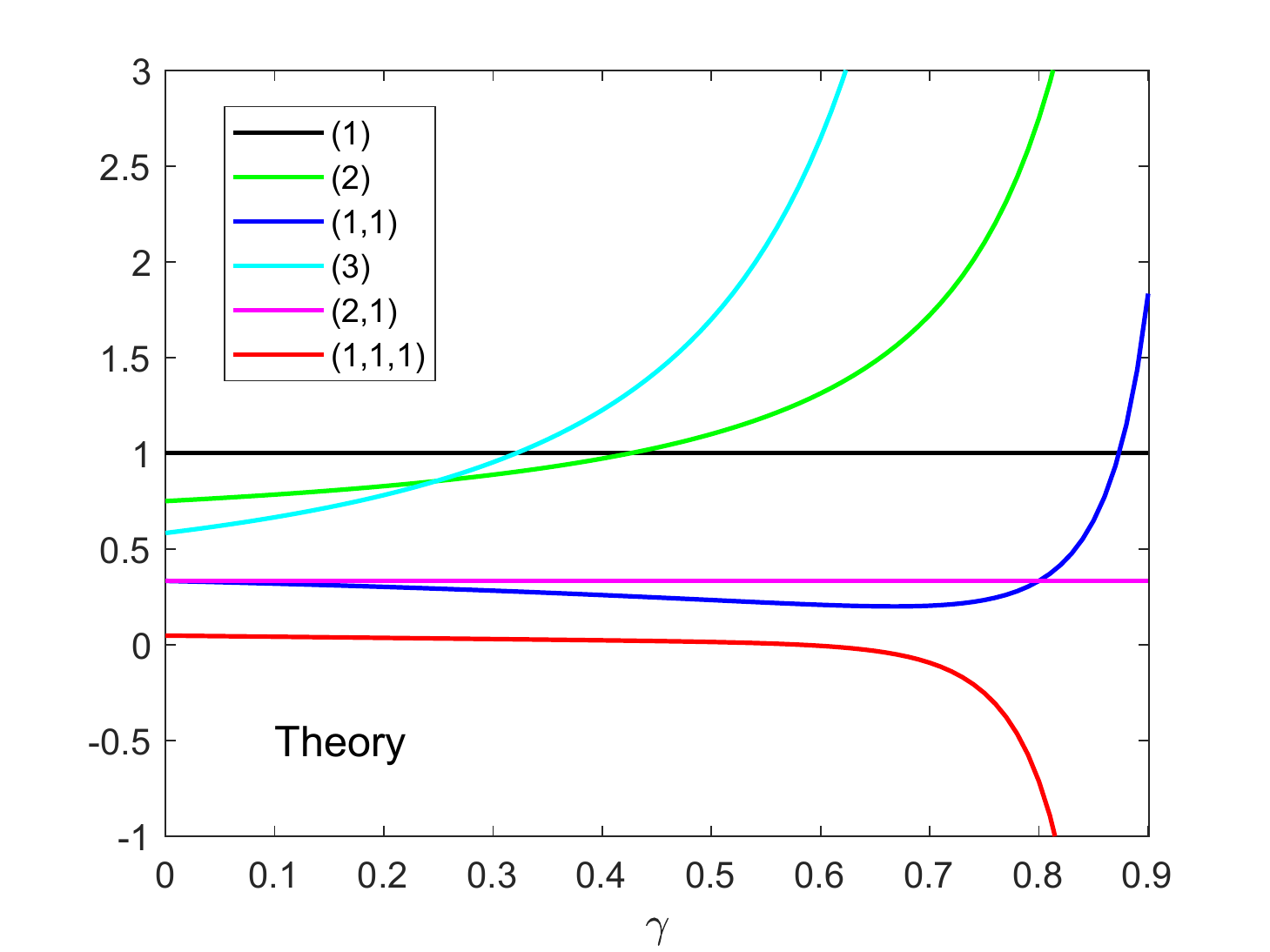}
\caption{Kicked rotor, $d=200$, $M=5$, first few moments $s_\lambda(Q)/(M\tau_D)^n$, averaged over energy and opening position. The vertical axis is truncated. Agreement is excellent, provided $\gamma$ is not too large. For $\gamma\gtrsim 0.7$ the average from the simulations can become unreliable because fluctuations are too strong.}
\end{figure*}

Finally, we computed $\langle p_{(n)}(Q)\rangle=\langle {\rm Tr}(Q^n)\rangle$ and $\langle p_{(1,...,1)}(Q)\rangle=M^n\langle \tau_W^n\rangle$, for the first few $n$, from the Schur-moments, according to (\ref{decomp}). The resulting exact expressions are convoluted, so we mention only limiting cases. To that end, let us define the transmission probability as $\Gamma=1-\gamma$.

When $M\gg 1$ and $M^{-2}\ll \Gamma\ll 1$, we are lead to conjecture that
\be\label{pn} \frac{1}{M}\langle p_{(n)}(Q)\rangle=\frac{(2n-2)!}{(n-1)!^2 \Gamma^{n-1}}\left(1+O\left(\frac{n^2}{M^{2}\Gamma}\right)\right)\ee for $n\ll M$ and $\langle p_{(n)}(Q)\rangle=\infty$ for $n>M$. On the other hand, when $\gamma$ is fixed and $M\Gamma\gg 1$, we conjecture that 
\be\label{p1n} \langle \tau_W^n\rangle(\gamma)=1+\frac{n(n-1)}{(M\Gamma)^2}+O((M\Gamma)^{-4})\ee  for $n\ll M$ and $\langle \tau_W^n\rangle=\infty$ for $n>M$. These are approximately the moments of a Gaussian distribution
\be \mathcal{P}_{\rm Wigner}(\tau)=\frac{1}{\sigma\sqrt{2\pi}}e^{-\frac{1}{2\sigma^2}(\tau-\tau_D)^2},\ee
with
\be \sigma=\frac{\sqrt{2}}{M\Gamma}.\ee We therefore expect this to be the distribution of the Wigner time delay in this regime. 

\section{Numerics for a specific system}\label{sec:numer}

Since we have presented some conjectures, we now check them in comparison to numerical results. We choose the traditional toy model of quantum maps.

Let $U$ be a unitary matrix, supposed to represent the quantum dynamics inside the cavity, as if it were closed. Its dimension, $d$, must be large in order to simulate the semiclassical limit. We take a rather modest $d=200$. From $U$, a scattering matrix $S_0$ of dimension $M$ can be computed by introducing coupling to the outside. The ratio $M/d$ must be small so that the open system is still reasonably similar to the closed one. We choose $M=5$.

The coupling is done by a $M\times d$ rectangular matrix $W$ as
\be S_0=-WUe^{i\epsilon}(1_d-PUe^{i\epsilon})^{-1}W^T,\ee
where $P=1_d-W^TW$ is a projector into the inside and $\epsilon$ plays the role of a quasi-energy. The interpretation of this formula is as follows: the quantum particle enters the cavity by means of $W^T$, then the geometric series $(1_d-e^{i\epsilon}PU)^{-1}$, which is like a Green's function, propagates it inside the cavity, and finally it exists by means of $W$.

The $S_0$ matrix corresponds to perfect coupling. Tunnel barriers are introduced by defining a futher transformation
\be S=-R+TS_0(1-RS_0)^{-1}T\ee 
where $R$ and $T$, representing reflection and transmission amplitudes of the barrier, are both multiples of the $M$-dimensional identity, $R^2=\gamma 1_M$ and $T^2=(1-\gamma)1_M$. The first term, $-R$, produces prompt reflection, while the second term is responsible for round trips of the particle inside the particle with multiple reflections at the barrier from the inside.

The Wigner-Smith matrix is then obtained as
\be Q=-i S^\dagger \frac{dS}{d\epsilon}.\ee The derivative is performed analytically, not numerically, keeping in mind that
\be \frac{d}{d\epsilon}(1-X)^{-1}=(1-X)^{-1}\frac{dX}{d\epsilon}(1-X)^{-1}.\ee

For the closed dynamics operator $U$ we have two choices, either sample it at random from the unitary group or take it to be a physical quantum map. We have checked that both approaches actually result in the same statistics, as is to be expected. 

We have used the kicked rotor, a conservative map on the torus. The equations of motion are
\be q_{n+1}=q_n+p_n,\ee
\be p_{n+1}=p_n+K\sin(2\pi q_n),\ee 
and the dynamics is known to be strongly chaotic if $K=9$, which is the value we use. This map is quantized by the matrix with elements given by $U_{jk}=\frac{1}{\sqrt{iN}}e^{i\Phi_{jk}},$
with 
\be \Phi_{jk}=\frac{\pi}{N}(j-k)^2-\frac{NK}{4\pi}\left(\cos(2\pi j/N)+\cos(2\pi k/N)\right).\ee

The way we generate statistics is by using $60$ different values for the quasi-energy $\epsilon$ and $20$ different positions for the opening, leading to $1200$ different matrices $Q$. The result is shown in Figure 2, where we plot different Schur-moments $\langle s_{\lambda}(Q)\rangle$ as functions of $\gamma$, with data from simulations on the left panel and our conjectured results on the right panel. 

We see that for the self-conjugated partitions $\lambda=(1)$ and $\lambda=(2,1)$ the corresponding Schur-moments indeed come out approximately independent of $\gamma$. The agreement is very good in all cases, so our conjectures are well validated. There are some discrepancies for large $\gamma$, but this is because in that regime there are wild fluctuations in the numerical results, so that much larger samples would be necessary in order to guarantee convergence of the average. The presence of factors $(1-\gamma)$ in the denominators of Schur-moments show that observables actually develop infinite variance in the limit $\gamma\to 1$, so perhaps such discrepancies are unavoidable.

We have also verified numerically that the estimates (\ref{pn}) and (\ref{p1n}) are indeed accurate (not shown).

\section{Conclusion}\label{sec:conc}

We have developed a semiclassical approach to the statistics of the time delay matrix for quantum systems with broken time-reversal symmetry and chaotic classical dynamics, in the presence of a tunnel barrier. The approach leads to results that are expressed as asymptotic series in powers of the reflectivity of the barrier, $\gamma$, with coefficients that are rational functions of the channel number, $M$. Based on calculations using computer algebra systems, we then conjectured some exact expressions for special kinds of Schur-moments, valid for arbitrary $\gamma$ and arbitrary $M$. These conjectures were then validated in comparison with numerical simulations.

This advance was made possible by combining the derivation of `efficient' diagrammatic rules like the ones from \cite{efficient} with the formulation in terms of matrix integrals proposed in \cite{matrix1,tunnel}. Together, these methods are able to go beyond even what a phenomenological random matrix theory is capable of delivering.

\end{document}